\documentclass[aps,prb,reprint,superscriptaddress]{revtex4-2}
\usepackage{epsfig}
\usepackage{dcolumn}
\usepackage{physics}
\usepackage{graphicx,graphics}
\usepackage{amsmath}
\usepackage{amssymb}
\usepackage{bm}
\usepackage{color}
\usepackage[utf8]{inputenc}
\usepackage[colorlinks,linkcolor={blue},citecolor={blue},urlcolor={blue}]{hyperref}
\usepackage{mathtools}
\usepackage{booktabs}
\usepackage{float}

\newcommand{\bk}{{\bm k}}

\begin{document}
  \title{Floquet-Driven Indirect Exchange Interaction Mediated by \\ Topological Insulator Surface States}

\author{Modi Ke}
\affiliation{Department of Physics and Astronomy, The University of Alabama, Tuscaloosa, AL 35487, USA}
\author{Mahmoud M. Asmar}
\affiliation{Department of Physics, Kennesaw State University, Marietta, Georgia 30060, USA}
\author{Wang-Kong Tse}
\affiliation{Department of Physics and Astronomy, The University of Alabama, Tuscaloosa, AL 35487, USA}

\begin{abstract}
  Light drives offer a potential tool for the dynamical control of magnetic interactions in matter. We theoretically investigate the indirect exchange coupling between two parallel chains of magnetic impurities on the surface of a topological insulator, driven by a time-periodic circularly polarized light field in the high-frequency, off-resonant regime. We derive a closed-form analytic expression for the spin susceptibility of the photon-dressed topological insulator surface states and obtain the irradiation dependence of the Ising, Heisenberg, and Dzyaloshinsky-Moriya exchange couplings between the impurity chains. Our results show a two-pronged modification of these exchange couplings by periodic drives. First, the RKKY oscillation period of the exchange couplings can be extended by enhancing the driving strength. Secondly, increasing driving strength enhances the envelope of RKKY oscillations of the Ising-type  while suppressing those of the Heisenberg-type and Dzyaloshinsky-Moriya-type. Our work provides useful insights for realizing Floquet engineering of collinear and non-collinear indirect exchange interactions in topological insulating systems.
\end{abstract}

\maketitle
\section{Introduction}\label{Intro}
Ruderman-Kittel-Kasuya-Yosida (RKKY) interaction \cite{RKKY1,RKKY2,RKKY3} is an indirect magnetic exchange coupling between localized magnetic moments mediated by the conduction electrons. The study of RKKY interaction has attracted significant interest because of its role as a long-range exchange interaction in magnetically doped systems. For host systems characterized by a parabolic band, the range function of the RKKY oscillations decays as $R^{-3}$ in three dimensions (3D) and as $R^{-2}$ in two dimensions (2D)~\cite{Kittel}. Typically, the RKKY interaction results in a parallel or antiparallel Heisenberg and/or Ising coupling between two magnetic impurities. However, the specific form of the RKKY interaction depends on the properties and band structure of the host materials~\cite{Narita, Flat}. In host materials characterized by a lack of inversion symmetry and considerable spin-orbit coupling, the Rashba effect plays an important role, inducing a twisted interaction between impurity spins~\cite{RKKYsoc2,RKKYsoc,twist} known as the Dzyaloshinskii-Moria (DM) interaction~\cite{Dzy,Moriya}. In addition, Rashba spin-orbit coupling enables control over the RKKY interaction through an external electric field~\cite{Zhu,helical,Nanoflake}. This presents a promising avenue for controlling the RKKY interaction in various materials and systems, enabling new opportunities for the development of advanced materials and devices with tunable magnetic properties. Furthermore, the relationship between the RKKY interaction, Rashba spin-orbit coupling, and external fields awaits further investigation to uncover the full extent of their interplay and potential applications.

The prototypical class of materials with a strong spin-orbit coupling is the 3D topological insulators (TIs) predicted in materials including $\rm{Bi_2Te_3}$, $\rm{Sb_2Te_3}$ and $\rm{Bi_2Se_3}$~\cite{Zhang}. The topological insulating state of the 3D TI consists of an insulating bulk spectrum separated by a band gap, where gapless surface states protected by time-reversal symmetry reside~\cite{topoprot}.
The RKKY interaction mediated by these surface states between individual impurity spins have been studied extensively~\cite{RKKYtopo,Parhi2,Mahmoud1}. In a TI dilutely doped with magnetic impurities on its surface, the RKKY interaction between the impurities is ferromagnetic when the chemical potential is at the Dirac point, favoring a ferromagnetic ordered state among the magnetic atoms \cite{Liu2}. When the chemical potential is away from the Dirac point, the RKKY interaction undergoes the typical Friedel oscillations changing sign between ferromagnetic and antiferromagnetic values \cite{Parhi1,Parhi2}. Due to strong Rashba spin-orbit coupling, the RKKY interaction in TIs couples not only the collinear components of the impurity spins but also the non-collinear components~\cite{RKKYtopo,Mahmoud1}. The exchange energy is characterized by collinear Heisenberg and Ising-like interactions as well as a non-collinear DM interaction, with the latter having a comparable magnitude as the former two~\cite{RKKYtopo,Mahmoud1,Parhi2}.

The interaction between topological matter and external
light drives has become a topic of vigorous research interest, including the dynamical
control and generation of topological phases from trivial ones by utilizing light
fields that are highly tunable and controllable. For instance, the periodic
driving of matter can induce a Floquet topological insulator state with chiral edge currents and other hallmark phenomena associated with topological phases in otherwise topologically trivial materials~\cite{Floqtopo,TopoFloq3,TopoFloq2,TopoFloq1}. Additionally, recent advancements in the field reveal that periodic driving can also give rise to higher-order Floquet topological phases. These phases are notable for their surface states, which, while remaining gapped in one lower dimension, exhibit unique wedge or corner modes localized at their lower-dimensional boundaries~\cite{HighTI1,HighTI2,HighTI3}. Light irradiation also enables the realization of Floquet Chern insulator states with both integer and fractional Chern numbers~\cite{FloqChern1,FloqChern2,FloqChern3,FloqChern4}. Beyond topological insulator states, it has been demonstrated that circularly polarized light can transform 3D Dirac materials into Floquet–Weyl semimetals, further expanding the scope of topological material states ~\cite{FloqWeyl1,FloqWeyl2,FloqWeyl3}. The spectrum of Floquet topological materials has also been enriched by the prediction of anomalous phases, such as anomalous Floquet-Anderson insulators~\cite{FloqAnd} and anomalous Floquet topological crystalline insulators~\cite{FloqTCI}, showcasing the diverse potential of periodic driving in manipulating material properties. The use of light sources to control materials' properties is not limited to topological phenomena, as it also allows dynamical control of effective electron-electron interactions, electronic hopping amplitudes, and lattice structures, which gives rise to a range of other phenomena such as light-induced superconductivity and metal-to-insulator phase transition~\cite{Sentef1}. 
Closely connected in the same vein is the concept of Floquet engineering of magnetic exchange interactions~\cite{Mentink2014,Mentink2015,Mentink2017,Bukov2016,Liu2018,Hejazi2019,Barbeau2019,Eckstein2017}. In particular, off-resonant optical driving has been demonstrated to provide a promising strategy for the dynamical coherent control of indirect exchange interactions mediated by conduction electrons in different irradiated materials such as graphene~\cite{Ke1,Ke2}, 2D magnetic lateral heterostructures ~\cite{Mahmoud2}, 3D magnetic vertical heterostructures~\cite{Mahmoud3}, and topological crystalline insulators~\cite{FloqTCI2}.

In this work, we theoretically study the RKKY interaction between two parallel impurity spin chains on the surface of a 3D TI driven out of equilibrium by CP (circularly polarized) light irradiation. We focus on the high-frequency, off-resonant regime and first obtain the static effective Hamiltonian based on the high-frequency expansion method~\cite{Vanvleck2,bukov}. The spin susceptibility tensor of the irradiated TI surface states is then derived analytically, which allows the indirect exchange interaction between the impurity spins mediated by the irradiated TI surface to be obtained.

Approximate analytical results of this exchange interaction energy provide the period and decay rate of the exchange interaction, which can be compared with the corresponding equilibrium results. Our numerical and analytical results reveal that both the period and the magnitude of the non-equilibrium RKKY oscillations are considerably modified by the driving field, stemming from light-induced effects on the electronic band structure and spin texture near the TI Fermi surface. Our discussion sheds light on the complex interplay between light-induced electronic perturbations and the RKKY interaction, and paves the way for further research into the magnetic properties in topological insulator systems under the influence of external driving fields.

This paper is organized as follows. First in Sec.~\ref{Formulation}, we introduce our model for the TI surface states under CP light irradiation and derive the corresponding Floquet Hamiltonian. We then provide a review of the derivation of the static effective Hamiltonian of the driven system in Sec.~\ref{EffHamiltonian} using the standard high-frequency expansion method. In Sec.~\ref{Susceptibility} we derive the spin susceptibility tensor of the irradiated topological insulator surface states. Next in Sec.~\ref{Exchange} we derive approximate analytical expressions for the RKKY exchange interaction as a function of driving strength and the separation between the two impurity spin chains. Finally, Sec.~\ref{Discussion} shows our numerical results performed using the developed formalism on the full effects of irradiation on the magnitude and period of exchange coupling oscillations and the comparison to analytical results, then we discuss the effects of CP light on the RKKY exchange interaction based on our results. We conclude our paper in Sec.~\ref{Conclusion} with a summary of our findings.

\section{Formulation}\label{Formulation}
We consider two parallel chains of magnetic impurities aligned in the $x$-direction and separated in the $y$-direction (Fig.~\ref{lattice}). The electronic Hamiltonian of the topological insulator surface states is given by the Dirac Hamiltonian \cite{RevmodernColloq,RevmodernTop,TopSurface},
\begin{eqnarray}\label{Hamilt1}
\mathcal{H}=\alpha_0 (\boldsymbol\sigma \times \boldsymbol{k})\cdot \hat z,
\end{eqnarray}
where $\alpha_0=0.328$ eV\AA\  is the Fermi velocity for Bi$_2$Se$_3$ and we denote the Fermi energy of the system to be $\mu$. A CP light field is normally incident on the topological insulator surface with a frequency $\Omega$ and a field amplitude  $E_{0}$, and the vector potential is given by ${\bm A}=E_0[\sin{(\Omega t)}\hat x-\cos{(\Omega t)}\hat y]/(\sqrt{2}\Omega)$. The CP light field couples to the Hamiltonian Eq.~\eqref{Hamilt1} via the minimal coupling scheme and the resulting time-dependent Hamiltonian of the irradiated system becomes
\begin{eqnarray}\label{Hamilt2}
  \mathcal{H}(t) &=& \alpha_0 (\sigma_x k_y-\sigma_y k_x)\nonumber
\\
&&-\alpha_0 \frac{e}{\hbar}A[\sigma_x \cos{(\Omega t)}+\sigma_y \sin{(\Omega t)}],
\end{eqnarray}
where $A= E_0/(\sqrt{2}\Omega)$ is the magnitude of the vector potential. We also define  $\mathcal{A}= e A \alpha_0/(\hbar^2\Omega)$ as the dimensionless driving strength.

As the Hamiltonian in Eq.~\eqref{Hamilt2} is time periodic, the Floquet-Bloch theorem is satisfied. The time periodic nature of the Floquet states,  $\left|u_{l,\bk}(t+T)\right\rangle=\left|u_{l,\bk}(t)\right\rangle$, where $T=2\pi/\Omega$ is the driving period, allows for a Fourier series representation of these states,
\begin{equation}
\left|u_{l,\bk}(t)\right\rangle=\sum_{n}e^{-i n \Omega t}\left|u_{l,\bk}^{n}\right\rangle,
\end{equation}
and
\begin{equation}\label{floqh}
\sum_n(\mathcal{H}_{mn}-n\hbar \Omega \delta_{m,n})|u_{l,\bk}^{n}\rangle=\epsilon_{l,\bk}|u_{l,\bk}^{m}\rangle,
\end{equation}
where $\mathcal{H}_{F,mn}=\mathcal{H}_{mn}-n\hbar \Omega \delta_{m,n}$ is the Floquet Hamiltonian \cite{Floq1,Floq2} and
\begin{equation}
\mathcal{H}_{mn}=\frac{1}{T}\int^{T}_{0}dt e^{i(m-n)\Omega t}\mathcal{H}(t),
\end{equation}
is the Floquet matrix.
Defining
\begin{eqnarray}\label{vvHamilt3}
&&\mathcal{H}_0=\alpha_0(\sigma_x k_y-\sigma_y k_x),~\mathcal{H}_1=-\alpha_0\frac{e}{\hbar}A\sigma_+,\nonumber\\
&&\mathcal{H}_{-1}=-\alpha_0\frac{e}{\hbar}A\sigma_-,~\mathcal{H}_{|m|>1}=0,
\end{eqnarray}
with $\sigma_{\pm}=(\sigma_x\pm i\sigma_y)/2$, the explicit form of our Floquet matrix can be written as
\begin{eqnarray}\label{flm}
\mathcal{H}_{mn}&=&\mathcal{H}_0\delta_{mn}+\mathcal{H}_{-1}\delta_{m+1,n}+\mathcal{H}_1\delta_{m-1,n}.
\end{eqnarray}
\section{Effective Hamiltonian}\label{EffHamiltonian}
\begin{figure}
   \begin{center}
            \includegraphics[width=1\columnwidth]{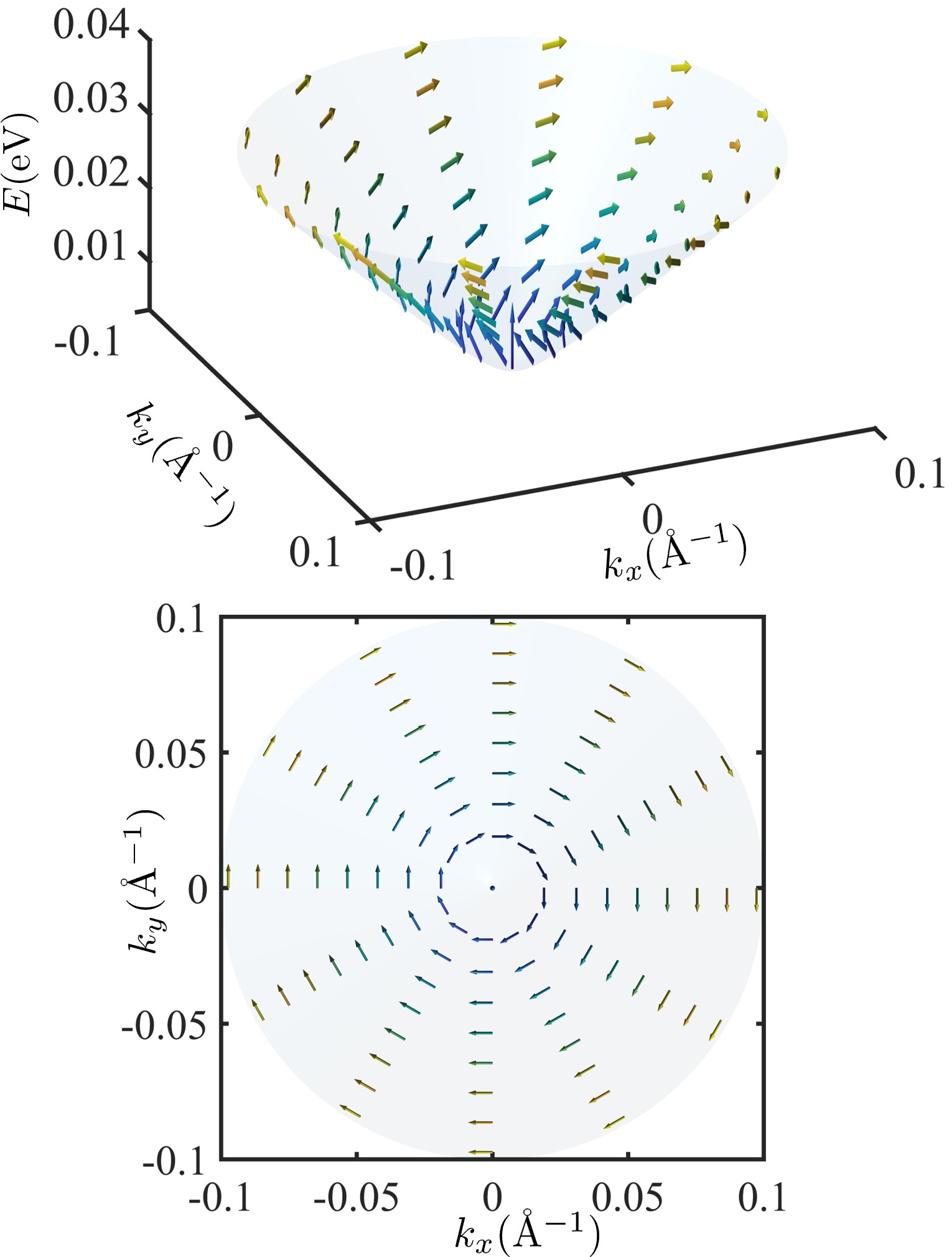}
                \end{center}
                \caption{Upper panel: conduction band and spin textures of an irradiated TI's surface states.
                Lower panel: projection of the spin texture on the $xy$-plane. The driving field is taken to have a frequency $\hbar\Omega=4.6$eV and strength $\mathcal{A}=0.13$.}\label{figspin1}

\end{figure}

\begin{figure}
   \begin{center}
            \includegraphics[width=1\columnwidth]{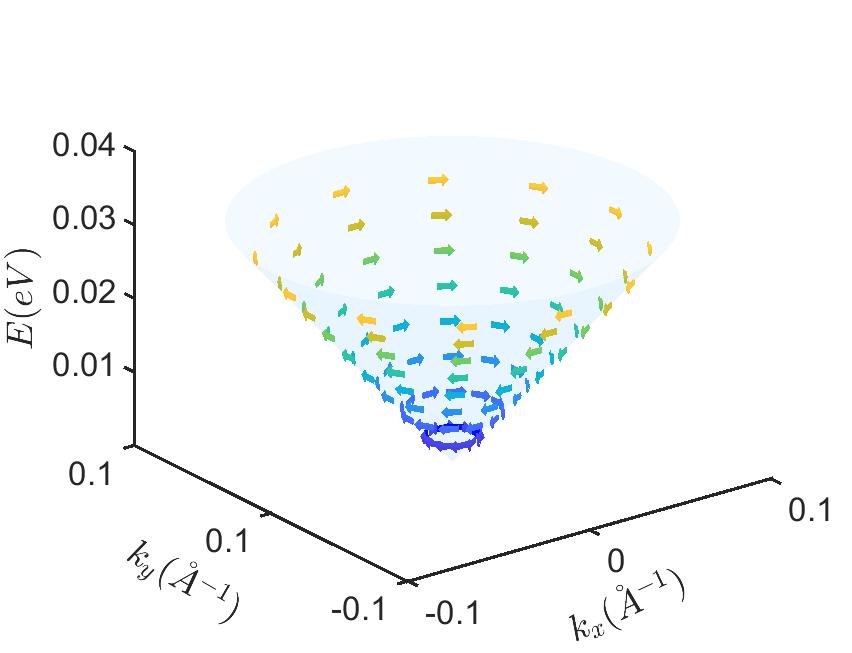}
                \end{center}
                \caption{Conduction band and spin textures of an TI's surface states in equilibrium. }\label{figspin2}

\end{figure}

In the regime of a high-frequency off-resonant light irradiation we can use the van Vleck perturbation theory to expand the full Floquet Hamiltonian matrix in powers of $1/\Omega$ into a finite-dimensional effective Hamiltonian \cite{Vanvleck2,bukov,Vanvleck1}. The van Vleck perturbation theory is effected by applying a unitary transformation in order to construct an effective Hamiltonian using the degenerate perturbation theory\cite{Vanvleck3,Vanvleck2}. The effective Hamiltonian is written in the form of the following series:
\begin{eqnarray}\label{vvHamilt1}
\mathcal{H}_{\rm eff}=\sum_{n=0}^{\infty} \mathcal{H}^{(n)}_{\rm vv}.
\end{eqnarray}
The leading terms of the expansion are given by
%
\begin{eqnarray}\label{vvHamilt2}
&&\mathcal{H}^{(0)}_{\rm vv}=0,\; \mathcal{H}^{(1)}_{\rm vv}=\mathcal{H}_0,\;\mathcal{H}^{(2)}_{\rm vv}=\sum_{m>0}\frac{[\mathcal{H}_{m},\mathcal{H}_{-m}]}{|m|\hbar \Omega}, \nonumber\\
&&\mathcal{H}^{(3)}_{\rm vv}=\\\nonumber
&&\sum_{m\neq0}\frac{[\mathcal{H}_{-m},[\mathcal{H}_{0},\mathcal{H}_{m}]]}{2(|m|\hbar \Omega)^2}+\sum_{m'\neq0,m}\frac{[\mathcal{H}_{-m'},[\mathcal{H}_{m'-m},\mathcal{H}_{m}]]}{3 m m'(\hbar \Omega)^2}.
\end{eqnarray}
For our system one gets the following effective Hamiltonian  up to second order in the high-frequency approximation which captures the effects of irradiation in the zero-photon regime~\cite{Eff1,Eff2,Eff3}:
\begin{eqnarray}\label{vvHamilt5}
\mathcal{H}_{\rm eff}=\alpha(\sigma_x k_y-\sigma_y k_x)+\Delta \sigma_z,
\end{eqnarray}
where we have $\alpha = \alpha_0[1-(e A \alpha_0/\hbar^2\Omega)^2]$ and $\Delta=(\alpha_0 e A/\hbar)^2/(\hbar \Omega)$. The above effective Hamiltonian takes the form of a single gapped Dirac Hamiltonian, with a light-induced band gap $\Delta$ and a renormalized Fermi velocity $\alpha_0 \to \alpha$.
Eq.~\eqref{vvHamilt5} yields the energy eigenvalues $E_{k,\eta}=\eta\sqrt{\alpha^2 k^2+\Delta^2}$ with the following eigenstates,
\begin{eqnarray}
&&|k,+\rangle=\begin{bmatrix}
 i \cos{\frac{\theta_k}{2}}\\
 \sin{\frac{\theta_k}{2}}e^{i\phi}
 \end{bmatrix}, |k,-\rangle=\begin{bmatrix}
 i \sin{\frac{\theta_k}{2}}\\
 -\cos{\frac{\theta_k}{2}}e^{i\phi}
 \end{bmatrix},
\end{eqnarray}
where $\eta=\pm$ labels the conduction band and valence band states, $\phi$ is the azimuthal angle of the momentum $k$, $\cos{\theta_k}=\Delta/\sqrt{\alpha^2 k^2+\Delta^2}$ and $\sin{\theta_k}=\alpha k/\sqrt{\alpha^2 k^2+\Delta^2}$.
Additional insights into the effects of light illumination can be obtained by calculating the expectation value of the electron spin associated with the conduction band electrons,
\begin{eqnarray}
\langle\sigma_x\rangle=\sin{\theta_k}\sin{\phi},\langle\sigma_y\rangle=-\sin{\theta_k}\cos{\phi},\langle\sigma_z\rangle=\cos{\theta_k}.\nonumber\\
\end{eqnarray}
Fig.~\ref{figspin1} depicts the calculated spin texture of the light driven system as a function of momentum and Fig.~\ref{figspin2} shows the equilibrium plot for comparison. It can be readily seen that near the band edge the spin predominantly points in the $z$-direction, while it tilts further and further away from the $z$-direction (\textit{i.e.}, lying closer and closer to the $xy$-plane) as one moves away from the band edge. This observation will be important in understanding the behavior of the RKKY interaction between impurity spins pointing in different directions, as we explain in Sec.~\ref{Discussion}.
\section{Spin Susceptibility}\label{Susceptibility}
\begin{figure}
   \begin{center}
            \includegraphics[width=1\columnwidth]{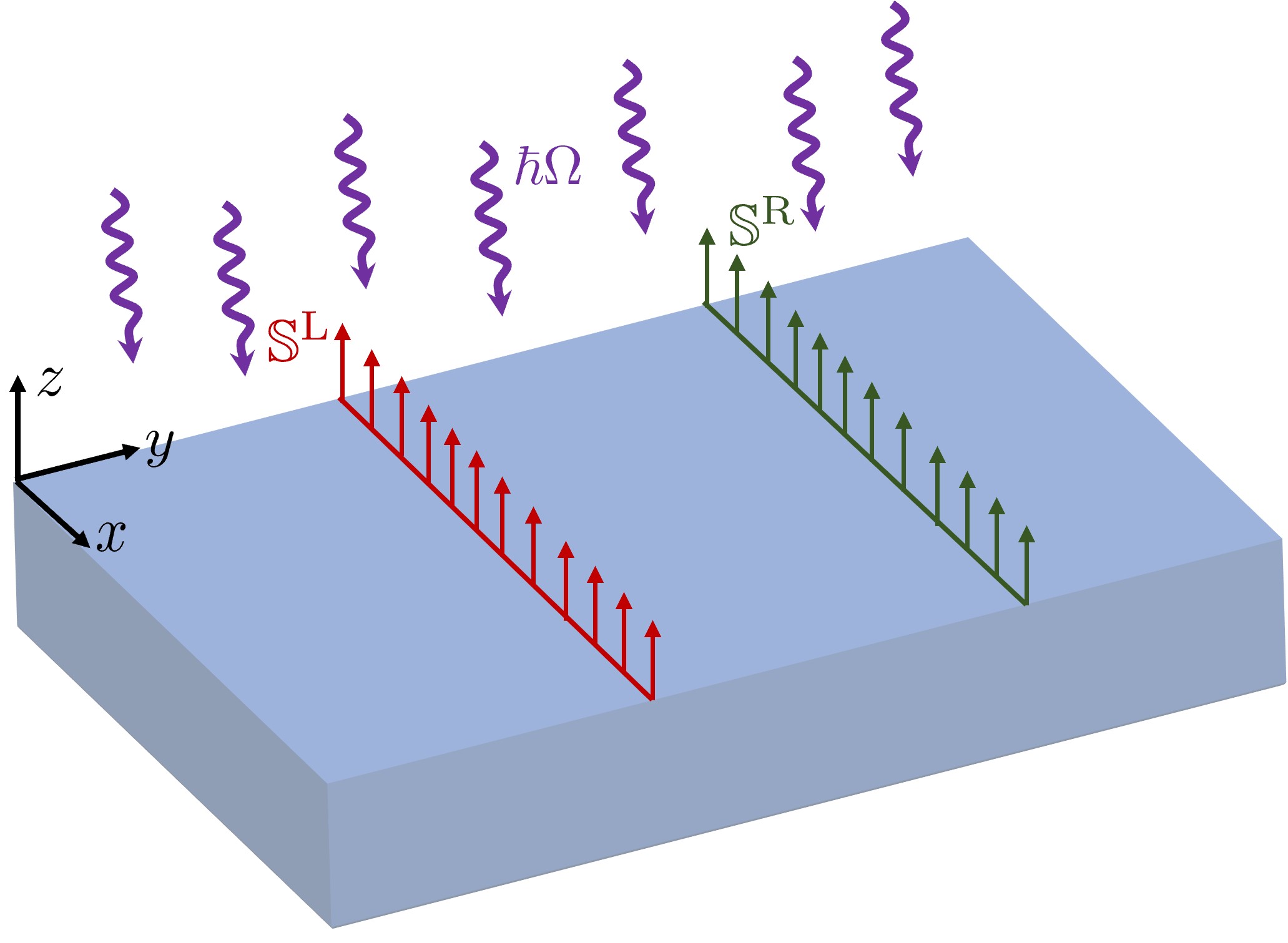}
                \end{center}
                \caption{Schematic of the system's setup. The topological insulator surface is deposited with two parallel chains of magnetic impurities. The driving light field is illuminated normally onto the topological insulator surface. }\label{lattice}
\end{figure}

In this paper we investigate the exchange interaction between two chains of magnetic impurities on the surface of a topological insulator irradiated by CP light. Our setup of two parallel impurity spin chains can be considered as the 2D analogue of the magnetic multilayers consisting of two ferromagnetic layers sandwiching a nonmagnetic spacer metal~\cite{Bruno1,Bruno2}. In our system (Fig.~\ref{lattice}), the two chains of magnetic impurities are assumed to consist of localized spins with magnetic moment $\mathbb{S}_i$ located at position $\boldsymbol{R}_i$ on the surface of the topological insulator. Each of the spins $\mathbb{S}_i$ is coupled to the electrons of the topological insulator surface states through the local potential  $V_i=J_0\delta(\boldsymbol{r}-\boldsymbol{R}_i)\boldsymbol{S}\cdot\mathbb{S}_i$. Here $J_0$ is the coupling strength and $\boldsymbol{S}$ is the magnetic moment of the electron spin of the topological insulator surface.
Our purpose is to calculate the exchange interaction energy between the two parallel chains of impurities on the surface of the topological insulator. Toward this end, we consider two impurities, one on the left chain (L) located at the origin and the other on the right chain (R) located at $\boldsymbol{R}_n$. 

The expectation value of the spin induced by the right-chain impurity is:
\begin{eqnarray}\label{spinchange}
  s^a=J_0 \sum_{b}\chi^{ab}(\boldsymbol{R}_n)\mathbb{S}^{\text{R}}_{b},
\end{eqnarray}
where $\chi^{ab}(\boldsymbol{R}_n)$ is the spin susceptibility or the spin correlation function, with the Matsubara spin susceptibility given by
\begin{eqnarray}\label{sus0}
\chi^{ab}(\boldsymbol{R}_n,\tau)=-\langle T_{\tau}\boldsymbol{S}^a(\boldsymbol{R}_n,\tau)\boldsymbol{S}^b(0,0)\rangle_0\;,
\end{eqnarray}
where $\tau$ is the imaginary time, $T_{\tau}$ is the imaginary time ordering operator and $\langle\cdots\rangle_0$ denotes the thermal average with respect to the effective Hamiltonian in Eq.~\eqref{vvHamilt5}. The exchange interaction energy between the two impurities is given by
\begin{eqnarray}\label{exchansingle}
E_n=J_0^2 \sum_{ab}\chi^{ab}(\boldsymbol{R}_n)\mathbb{S}^{\text{L}}_{a}\mathbb{S}^{\text{R}}_{b}.
\end{eqnarray}

In order to determine the exchange energy, we first proceed to calculate the spin susceptibility in this section. Using Wick's theorem, Eq.~\eqref{sus0} can be written in terms of the single-particle Green's functions $G$ as
\begin{eqnarray}\label{sus1}
&&\chi^{ab}(\boldsymbol{q},iq_n)=\\
&&\frac{\mu_B^2}{\beta}\sum_{i k_n}\int\frac{d^2 k}{(2\pi)^2}{\rm Tr}\{G(\boldsymbol k+\boldsymbol q,i k_n+ i q_n)\sigma_aG(\boldsymbol k,i k_n)\sigma_b\},\nonumber
\end{eqnarray}
where $\mu_B$ is the Bohr magneton, $\beta=1/(k_B T)$ and the single-particle Matsubara Green’s function is given by
\begin{eqnarray}\label{Green}
&&G(\boldsymbol{k},i k_n)=(i k_n-\mathcal{H}_{\rm eff})^{-1}\nonumber\\
&&=\frac{(i k_n +\mu)+\alpha(\boldsymbol{\sigma}\times\boldsymbol{k})_z+\Delta\sigma_z}{(i k_n +\mu)^2-(\alpha^2 k^2+\Delta^2)}.
\end{eqnarray}
The denominator of Eq.~\eqref{sus1} can be shown to have the following form using the Feynman parameterization~\cite{Liu1}:
\begin{eqnarray}\label{Denom}
&&\frac{1}{[(ik_n+\mu)^2-E_{\boldsymbol{k}+\boldsymbol{q}}^2][(ik_n+\mu)^2-E_{\boldsymbol{k}}^2]}\\
&&=\int^{1}_{0}dx\frac{1}{[\alpha^2(\boldsymbol{k}+x\boldsymbol{q})^2+(k_n-i\mu)+\Delta^2+\delta^2(x)]^2},\nonumber
\end{eqnarray}
where $x$ here is an auxiliary parameter to help in evaluating the integral and $\delta(x)=x(1-x)\alpha^2 q^2$.

In the zero-temperature limit, the Matsubara sum in Eq.~\eqref{sus1} turns into an integral
and the susceptibility can be written in the following form,
\begin{eqnarray}\label{sus2}
&&\chi^{ab}=\mu_B^2\int^1_0 dx\int\frac{dk_0}{(2\pi)}\int\frac{d^2 k}{(2\pi)^2}\nonumber\\
&&\frac{1}{[\alpha^2 k^2+(k_0-i \mu)^2+\Delta^2+\delta^2(x)]^2}\nonumber\\
&&\times{\rm Tr}\{[(i k_0+\mu)+[\alpha(\boldsymbol k+(1-x)\boldsymbol q)\times\boldsymbol\sigma]_z+\Delta \sigma_z]\sigma_a\nonumber\\
&&(i k_0+\mu +[\alpha (\boldsymbol k-x\boldsymbol q)\times\boldsymbol\sigma]_z+\Delta \sigma_z)\sigma_b\},
\end{eqnarray}
where $k_0$ plays the role of $k_n$ after turning the Matsubara sum into an integral.

We calculate the components of the susceptibility tensor using the above equation (details of the calculation are relegated into Appendix~\ref{app1}). The susceptibility tensor is then found to be
\begin{eqnarray}\label{sus3}
&&\chi=\mu_B^2\begin{bmatrix}
 g_1 \cos^2{\phi}&\frac{1}{2}g_1\sin{2\phi}&-i g_2 \cos{\phi}\\
 \frac{1}{2}g_1\sin{2\phi}& g_1 \sin^2{\phi}&-i g_2 \sin{\phi}\\
 i g_2 \cos{\phi}&i g_2 \sin{\phi}& g_3
 \end{bmatrix},
\end{eqnarray}
where we have defined the following piecewise functions with $\gamma=\sqrt{1-4\mu^2/(\alpha^2 q^2+4\Delta^2)}$:
\begin{eqnarray}\label{sus4}
&&g_1=\nonumber\\
&&\begin{cases}
          \frac{1}{4\pi\alpha^2}\Re\bigg[\sqrt{1-4\frac{\mu^2-\Delta^2}{\alpha^2 q^2}}|\mu|+\frac{1}{2}(\alpha q-\frac{4\Delta^2}{\alpha q})\sin^{-1}\gamma\bigg],&\\ \text{if } |\mu|> \Delta
          \\
        \frac{1}{2\pi\alpha^2}\bigg[\frac{1}{2}\Delta+\alpha q \tan^{-1}{\frac{\alpha q}{2\Delta}}\\
        -\frac{1}{8}(3\alpha q+\frac{4\Delta^2}{\alpha q})(\pi-\sin^{-1}\gamma)\bigg], & \\
        \text{if }  |\mu|< \Delta
    \end{cases}
\end{eqnarray}
\begin{equation}\label{sus5}
g_2=
\begin{cases}
          -\frac{q}{4\pi\alpha}\bigg\{1-\Re\bigg[\sqrt{1-4\frac{\mu^2-\Delta^2}{\alpha^2 q^2}}\bigg]\bigg\},& \text{if } |\mu|> \Delta\\
          \\
       0,& \text{if }  |\mu|< \Delta
    \end{cases}
\end{equation}
\begin{equation}\label{sus6}
g_3=
\begin{cases}
          \frac{1}{2\pi\alpha^2}\Re\bigg[|\mu|-2\Delta+\frac{1}{2}(\alpha q+\frac{4\Delta^2}{\alpha q})\sin^{-1}\gamma\bigg],& \\
          \text{if } |\mu|> \Delta\\
          \\
        \frac{\alpha q+\frac{4\Delta^2}{\alpha q}}{4\pi\alpha^2}\Re\bigg[\sin^{-1}\gamma\bigg], & \\
        \text{if }  |\mu|< \Delta
    \end{cases}
\end{equation}
It is worthwhile to note that the general angular dependence for each term in Eq.~\eqref{sus3} can also be deduced from symmetry arguments~\cite{symmetry}. 

\section{Exchange interaction between magnetic impurity chains}\label{Exchange}

We now consider the exchange interaction energy per unit length of the two parallel chains of impurities along the $x$-direction of the topological insulator plane separated by distance $y$. We use Eq.~\eqref{exchansingle} and Eq.~\eqref{sus3} to derive the exchange interaction energy between the two impurity chains in the momentum space representation. This representation is convenient as it simplifies the sum over the impurities along the impurity chain. In our set up we assume the impurities of a chain on the surface to be located on the lattice sites of Bi atoms with each of the neighbouring impurities separated by two Se atoms, and then the distance between two adjacent impurities on one chain is $2A_0/c$, where $A_0$ is the area of the unit cell and $c=4.19$\AA\ is the distance between Bi atoms on the surface. We first calculate the interaction between the two impurity chains and then divide by the length of the chain to get the averaged exchange interaction per unit length of the impurity chains. Then the exchange coupling per unit length is given by~\cite{Mahmoud1} :
\begin{eqnarray}\label{exch1}
I=\sum_{a,b}\frac{J_0^2 c}{(2\pi)^2 2A_0}\mathbb{S}^{\text{L}}_a \mathbb{S}^{\text{R}}_b \int d \boldsymbol q \chi_{ab}(\boldsymbol q) \sum_{n}e^{i \boldsymbol q \cdot \boldsymbol R_n},
\end{eqnarray}
where in the last sum we sum over all the impurities along one chain. With the periodic boundary condition in the $x$-direction the sum $\sum e^{i q_x x}$ is nonzero only for $q_x=0$, and thus the exchange coupling can be written as
\begin{eqnarray}\label{exch2}
I(y)=\sum_{a,b}\frac{J_0^2 c^2}{2\pi (2A_0)^2}\mathbb{S}^{\text{L}}_a \mathbb{S}^{\text{R}}_b \int dq_y \chi_{ab}(q_x=0,q_y) e^{i q_y y},\nonumber\\
\end{eqnarray}
where we used the length of 1D Brillouin zone $(2\pi c)/A_0$.
In our case, we have $q_x=0$ and $\phi=\pm\pi/2$, so the spin susceptibility tensor reduces to
\begin{eqnarray}\label{redsus}
&&\chi=\mu_B^2\begin{bmatrix}
 0&0&0\\
 0& g_1 &-i g_2 {\rm sgn}(q_y)\\
 0&i g_2 {\rm sgn}(q_y)& g_3
 \end{bmatrix}.
\end{eqnarray}
Thus we have:
\begin{eqnarray}\label{intE}
I(y)=A(y) \mathbb{S}^{\text{L}}_y \mathbb{S}^{\text{R}}_y + B(y) \mathbb{S}^{\text{L}}_z \mathbb{S}^\text{{R}}_z + D(y) (\mathbb{S}^{\text{L}}\times\mathbb{S}^{\text{R}})\cdot\hat{x},
\end{eqnarray}
where the $A$, $B$ and $D$ terms determine the real space dependence of the Heisenberg, Ising and Dzyaloshinsky-Moriya exchange interactions respectively. Among the three terms the antisymmetric DM exchange interaction is attributed to the effect of the Rashba spin-orbit coupling characterizing TI's surface states~\cite{DM1,DM2,DM3}. Therefore we need to calculate the following integrals:
\begin{eqnarray}\label{fourier1}
A(y)=\frac{1}{8\pi}\left(\frac{J_0 c}{A_0}\right)^2 \int_{-\infty}^{\infty}dq_y e^{i q_y y}\chi_{yy}(q_y),\nonumber\\
B(y)=\frac{1}{8\pi}\left(\frac{J_0 c}{A_0}\right)^2 \int_{-\infty}^{\infty}dq_y e^{i q_y y}\chi_{zz}(q_y),\\
D(y)=\frac{1}{8\pi}\left(\frac{J_0 c}{A_0}\right)^2 \int_{-\infty}^{\infty}dq_y e^{i q_y y}\chi_{yz}(q_y).\nonumber
\end{eqnarray}
In this work, we are interested in the case of a finite Fermi level away from the light-induced gap. These integrals can be evaluated analytically for $\vert \mu\vert >\Delta$ by approximating the integral around the Kohn anomaly (\textit{i.e.} a Fermi surface singularity)~\cite{Kohn}. 

We evaluate the integral around a small neighborhood of the Kohn anomaly at $q_0=2\sqrt{\mu^2-\Delta^2}/\alpha$ and assume that the change of the integrand is dominated by the $q-q_0$ term, treating the other terms as constants evaluated at $q=q_0$. Within this approximation we arrive at the following asymptotic forms for the interaction energy in Eq.~\eqref{intE} as $I=I_A+I_B+I_D$:

\begin{eqnarray}\label{analyticalA}
&&I_A(y)\approx-I_{A,0}2\sqrt{\pi}\sqrt{\frac{c}{\alpha_0}}\frac{1}{|\mu|}\bigg(\frac{\sqrt{\mu^2-\Delta^2}}{1-\frac{\Delta}{\hbar\Omega}}\bigg)^{\frac{3}{2}}\nonumber\\
&&\times\bigg(\frac{c}{y}\bigg)^{\frac{3}{2}}\cos{\left(q_0 y -\frac{\pi}{4}\right)},
\end{eqnarray}
\begin{eqnarray}\label{analyticalB}
&&I_B(y)\approx-I_{B,0}2\sqrt{\pi}\sqrt{\frac{c}{\alpha_0}}\frac{|\mu|}{\mu^2-\Delta^2}\bigg(\frac{\sqrt{\mu^2-\Delta^2}}{1-\frac{\Delta}{\hbar\Omega}}\bigg)^{\frac{3}{2}}\nonumber\\
&&\times\bigg(\frac{c}{y}\bigg)^{\frac{3}{2}}\cos{\left(q_0 y -\frac{\pi}{4}\right)}
\end{eqnarray}
\begin{eqnarray}\label{analyticalD}
&&I_D(y)\approx-I_{D,0}2\sqrt{\pi}\sqrt{\frac{c}{\alpha_0}}\frac{1}{\sqrt{\mu^2-\Delta^2}}\bigg(\frac{\sqrt{\mu^2-\Delta^2}}{1-\frac{\Delta}{\hbar\Omega}}\bigg)^{\frac{3}{2}}\nonumber\\
&&\times\bigg(\frac{c}{y}\bigg)^{\frac{3}{2}}\cos{\left(q_0 y -\frac{3\pi}{4}\right)},
\end{eqnarray}

where we have used $I_{A,0}=\mathcal{I}\mathbb{S}^{\text{L}}_y\mathbb{S}^{\text{R}}_y$, $ I_{B,0}=\mathcal{I}\mathbb{S}^{\text{L}}_z\mathbb{S}^{\text{R}}_z$, $I_{D,0}=\mathcal{I}(\mathbb{S}^{\text{L}}\times\mathbb{S}^{\text{R}})_x$, and $\mathcal{I}= (\mu_BJ_0/(4\sqrt{2\alpha_0}\pi A_0))^2$. As are Eqs.~\eqref{sus4}-\eqref{sus6}, Eqs.~\eqref{analyticalA}-\eqref{analyticalD} are independent of the signs of $\mu$ due to the particle-hole symmetry of the Dirac Hamiltonian. In equilibrium, the interaction for impurity spins along the $z$-direction $I_A$ and that along the $y$-direction $I_B$ are precisely identical to each other. However, with irradiation it can be seen that the two terms become distinct. In Appendix \ref{app3}, we provide an alternative derivation of Eqs.~\eqref{analyticalA}-\eqref{analyticalD} above using semi-classical real-space Green's functions obtained from the method of stationary phase approximation.

\begin{figure}

   \begin{center}
            \includegraphics[width=1\columnwidth]{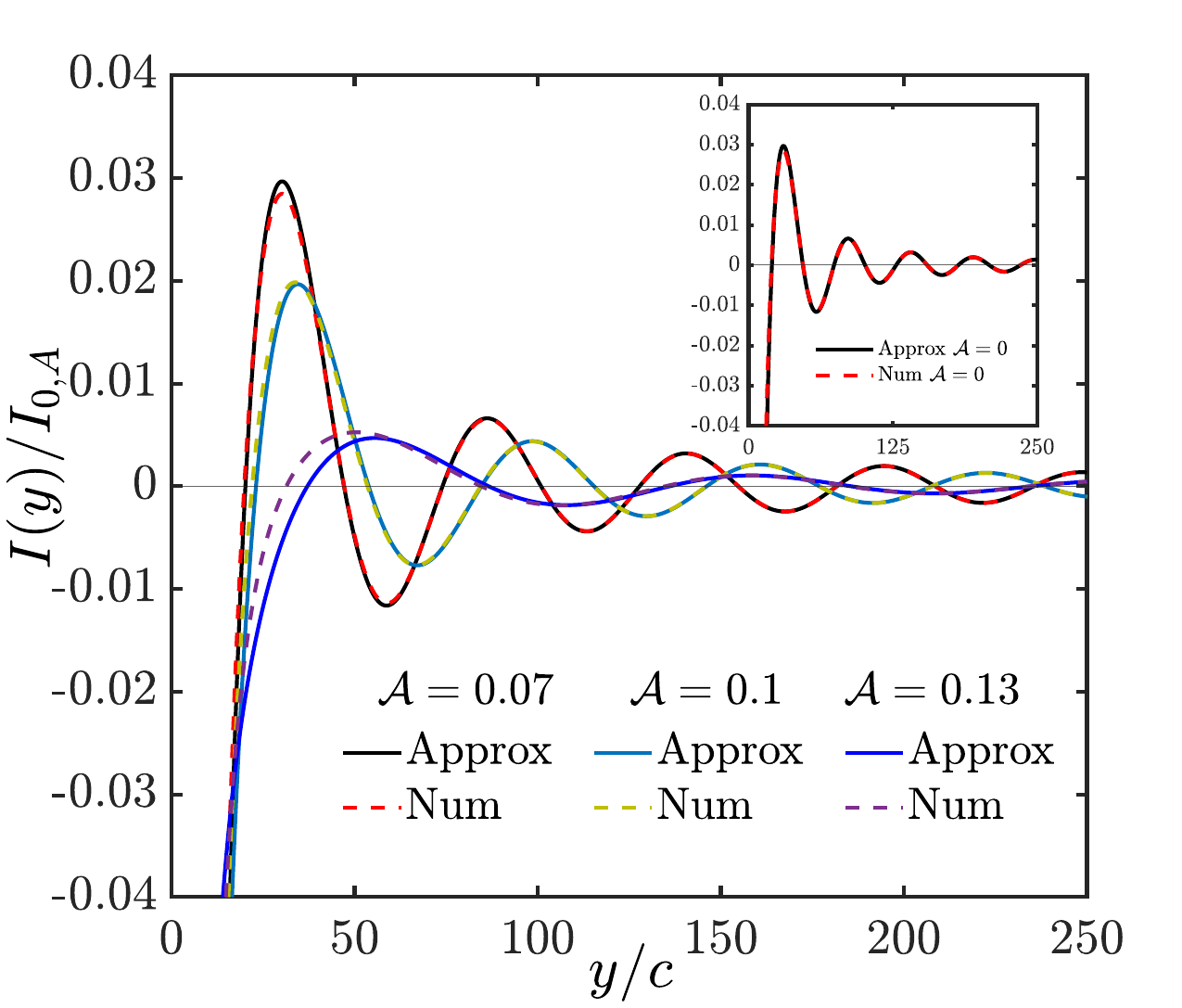}
                \end{center}
                \caption{The $yy$ component of the exchange coupling $I_A(y)/I_{A,0}$ for different driving strengths $\mathcal{A}$ as a function of the inter-chain separation. Numerical results are obtained from Eq.~\eqref{fourier1} and approximated analytical result from Eq.~\eqref{analyticalA}. Here $c=4.19$\AA, the Fermi energy $ \mu=0.092 $eV, and the driving field is taken to have a frequency $\hbar\Omega=4.6$eV.
                }\label{fig1}

\end{figure}

Although our primary interest in this work is in spin chains, for completeness and the purpose of comparison, we have also calculated the long-range asymptotes of the RKKY interaction energy between two isolated impurity spins on an illuminated TI surface (see Appendix \ref{app3}). This allows us to clearly delineate the qualitative effects due to irradiation from those due to the spin chains. Comparing the long-range behaviors for the two cases [Eqs.~\eqref{SPASingleA}-\eqref{SPASingleD} and Eqs.~\eqref{analyticalA}-\eqref{analyticalD}] reveals three key observations. In going from isolated impurities to spin chains, (1). the power-law decay becomes slower, changing from $y^{-2}$ to $y^{-3/2}$; (2). the oscillation phase acquires an additional $\pi/4$; (3). the oscillation amplitudes remain the same up to an overall multiplicative factor that depends, among other parameters, on both $\Delta$ and $\Omega$.


\section{Discussion}\label{Discussion}
\begin{figure}

   \begin{center}
            \includegraphics[width=1\columnwidth]{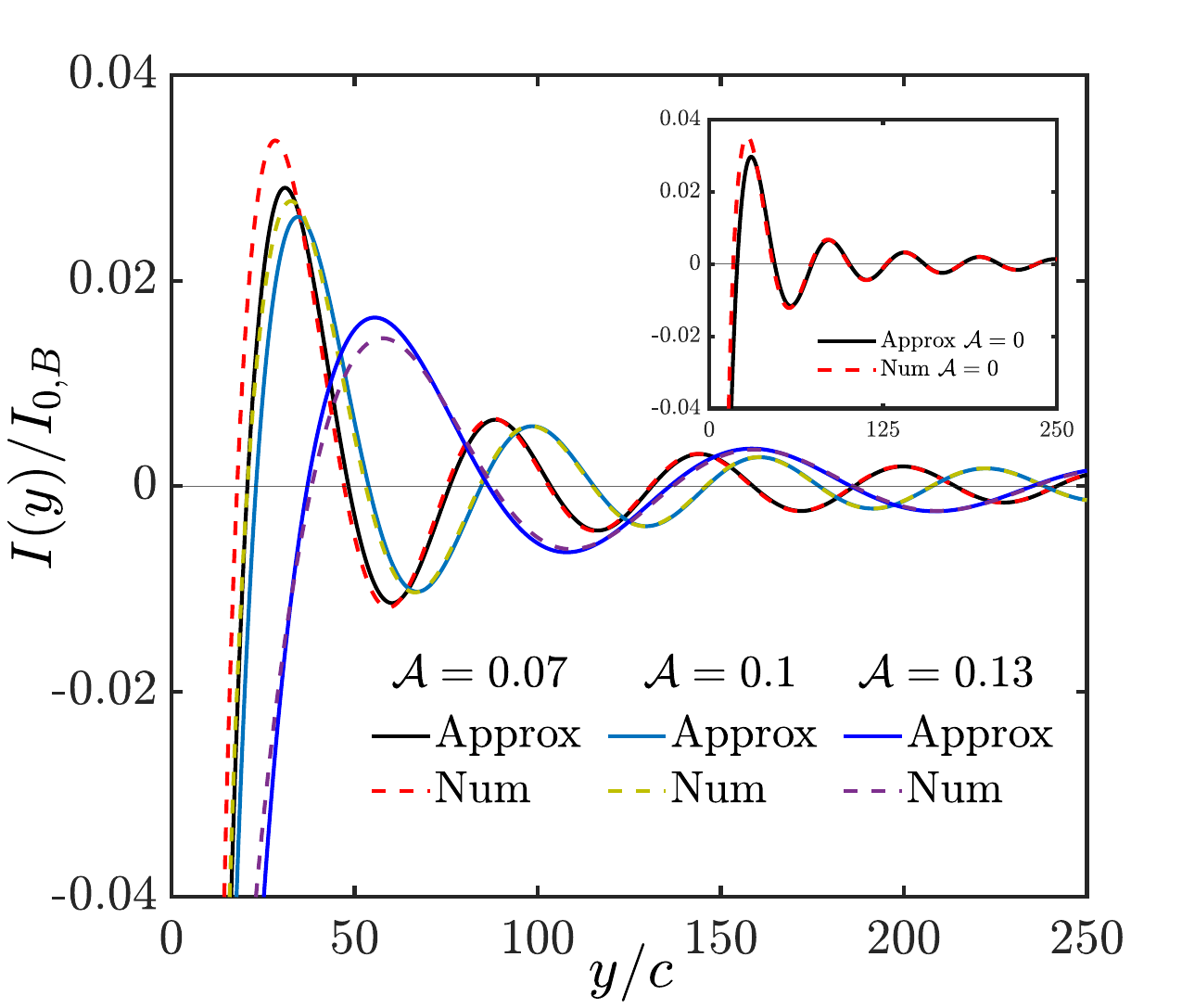}
                \end{center}
                \caption{The $zz$ exchange coupling component of $I_B(y)/I_{B,0}$ as a function of the inter-chain separation for different driving strengths $\mathcal{A}$. The numerical integration of Eq.~\eqref{fourier1} results into the displayed numerical results while Eq.~\eqref{analyticalA} gives the approximated analytical results.
                The values of $c$, $ \mu$, and $\hbar\Omega$ are given in Fig.~\ref{fig1}.}\label{fig2}

\end{figure}

\begin{figure}

   \begin{center}
            \includegraphics[width=1\columnwidth]{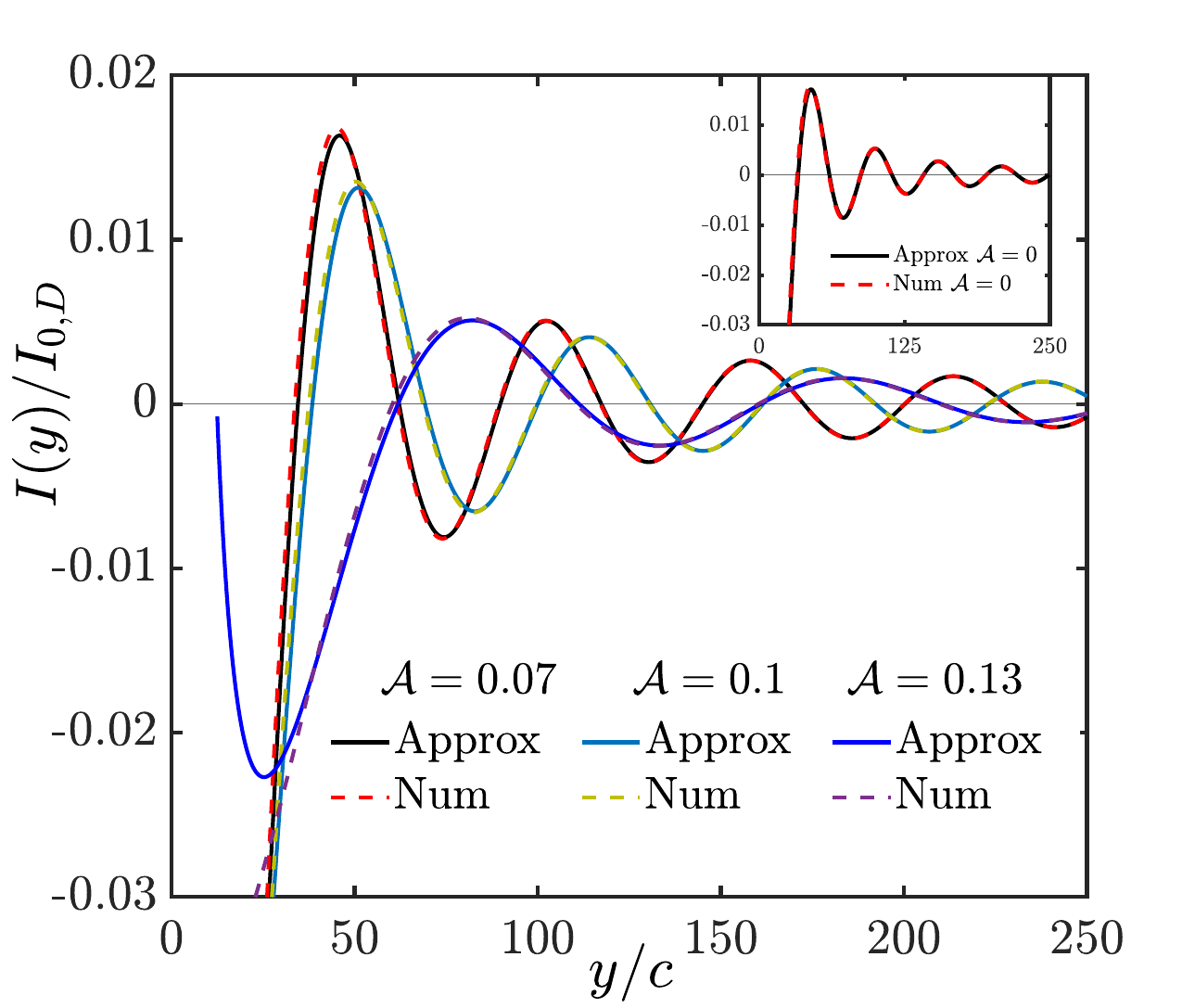}
                \end{center}
                \caption{The non-collinear, $yz$ and $zy$, components of the exchange coupling $I_D(y)/I_{D,0}$ for different values of $\mathcal{A}$ as a function of $y$, the inter-chain separation, for the values of $\mu$, $\hbar\Omega$ and $c$ used in Fig.~\ref{fig1}. The approximate analytical result are obtained from Eq.~\eqref{analyticalA}, and the numerical ones from Eq.~\eqref{fourier1}.}\label{fig3}
\end{figure}

A comparison of the approximate analytical results in Eqs.~\eqref{analyticalA}-~\eqref{analyticalD} with the numerical results directly calculated from  Eq.~\eqref{fourier1} is shown in Figs.~\ref{fig1}-\ref{fig3} for the three components and for different driving strengths. The agreement between the numerics and the analytical approximation is already very close at distances around $y=20-30c$ and the two results overlap at larger distances.


\textit{Decay power law.} 
Eqs.~\eqref{analyticalA}-~\eqref{analyticalD} show that irradiation does not change the decay power law, and the envelope still decays like $y^{-3/2}$ as in equilibrium. This follows from the fact that the decay power law only depends on the energy spectrum of the conduction electrons mediating the RKKY interaction, the dimensionality of the spacer material, and the dimensionality of the impurity spin configuration. 

\textit{Phase constant.} Eqs.~\eqref{analyticalA}-~\eqref{analyticalD} demonstrate that the phase relationship among the three components remains unaltered by light irradiation, suggesting that light irradiation (at least in the off-resonant, high-frequency regime considered in this work) does not induce an additional phase that would otherwise modify the relative strength of each term at a fixed distance.

\textit{Oscillation period.} The Heisenberg (yy),  Ising (zz), and DM (yz and zy) exchange interactions are characterized by the same oscillation period, given from the long-range analytic results Eqs.~\eqref{analyticalA}-~\eqref{analyticalD} as $2\pi/q_0$. Since the Kohn anomaly wavevector $q_0$ for a fixed Fermi level is determined by the band edge of the light-induced gap, one can expect the oscillation period to vary with the light intensity. Figs.~\eqref{fig1}-~\eqref{fig3} show that all three exchange interaction terms oscillate more slowly with distance at a larger value of driving strength $\mathcal{A}$. When the driving field is increased, the band edge is pushed closer to the Fermi level resulting in a smaller $q_0$, leading to a longer oscillation period.

\begin{figure}\label{mag}

   \begin{center}
            \includegraphics[width=1\columnwidth]{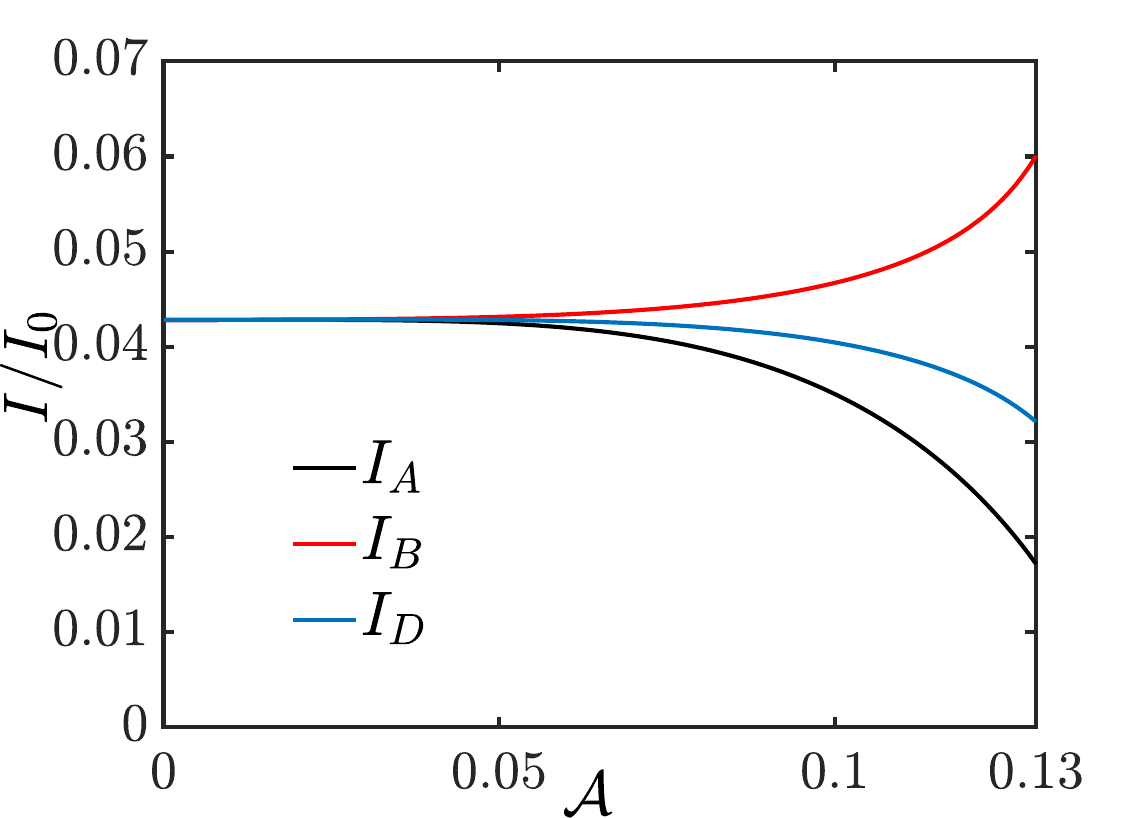}
                \end{center}
                \caption{Plot of the oscillation envelope of the exchange coupling $I/I_{0}$ as a function of the driving strength $\mathcal{A}$, obtained from the analytical results Eq.~\eqref{analyticalA}, \eqref{analyticalB}, and \eqref{analyticalD}. The driving field is taken to have a frequency $\hbar\Omega=4.6$eV, the inter-chain separation is fixed at $y=10c$ and the Fermi energy is $ \mu=0.092 $eV.}\label{fig4}

\end{figure}

\textit{Oscillation envelope.} 
We now examine the amplitude of the oscillations. 
Fig.~\ref{fig4} shows the envelopes of $I_{A}$, $I_{B}$ and $I_{D}$ as a function of the driving strength $\mathcal{A}$. We first notice that the envelopes of $I_{A}$, $I_{B}$ and $I_{D}$ are all equal in equilibrium, as can be seen by taking $\Delta=0$ and $\alpha = \alpha_0$ in Eqs.~\eqref{analyticalA}-~\eqref{analyticalD}. When the driving field is turned on, the three exchange coupling terms are modified differently. As seen in Fig.~\ref{fig4}, increasing the driving strength $\mathcal{A}$ decreases the oscillation envelopes for the $yy$ and $yz$ terms but increases the envelope for the $zz$ term. This can also be seen by expanding the oscillation envelopes in the long-range analytic results Eqs.~\eqref{analyticalA}-~\eqref{analyticalD} in powers of the driving strength $\mathcal{A}=e A \alpha_0/(\hbar^2\Omega)$:

\small
\begin{eqnarray}\label{Taylor}
I_A\propto I_{A,0}\sqrt{\frac{\pi|\mu| c}{\alpha_0}}\left(\frac{c}{y}\right)^\frac{3}{2}\bigg[2+3\mathcal{A}^2-\frac{3}{4}\left(2\frac{\hbar^2\Omega^2}{\mu^2}-5\right)\mathcal{A}^4\cdots\bigg],\nonumber\\
I_B\propto I_{B,0}\sqrt{\frac{\pi|\mu| c}{\alpha_0}}\left(\frac{c}{y}\right)^\frac{3}{2}\bigg[2+3\mathcal{A}^2+\frac{1}{4}\left(2\frac{\hbar^2\Omega^2}{\mu^2}+15\right)\mathcal{A}^4\cdots\bigg],\nonumber\\
I_D\propto I_{D,0}\sqrt{\frac{\pi|\mu| c}{\alpha_0}}\left(\frac{c}{y}\right)^\frac{3}{2}\bigg[2+3\mathcal{A}^2-\frac{3}{4}\left(2\frac{\hbar^2\Omega^2}{\mu^2}-15\right)\mathcal{A}^4\cdots\bigg].\nonumber\\
\end{eqnarray}
\normalsize
Eq.~(\ref{Taylor}) shows that while the leading zeroth order and second order terms are the same for all three components of the exchange energy, the dominant fourth order term is positive for the $zz$ component and negative for the $yy$ and $yz$ components, respectively.

\begin{figure}\label{mag2}

   \begin{center}
            \includegraphics[width=1\columnwidth]{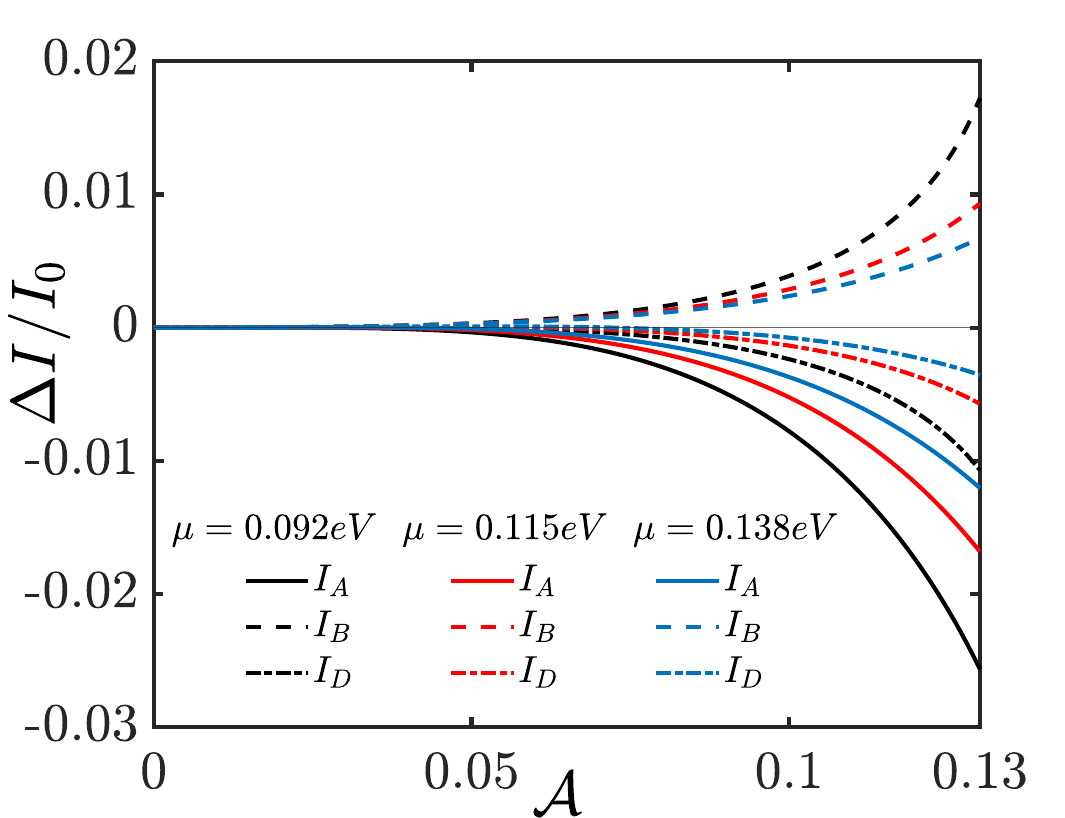}
                \end{center}
                \caption{The oscillation envelope of the exchange coupling measured from its equilibrium value, $\Delta I=I-I(\mathcal{A}=0)$, as a function of the driving strength $\mathcal{A}$,  obtained from Eq.~\eqref{analyticalA}, \eqref{analyticalB}, and \eqref{analyticalD}. The driving field is taken to have a frequency $\hbar\Omega=4.6$eV and  the inter-chain separation is fixed at $y=10c$. Results for several different Fermi energies $ \mu=0.092,0.115,0.138 $eV are shown.}\label{fig5}

\end{figure}

One can understand the above observation from the light-induced spin texture change of the topological insulator surface states as shown in Fig.~\ref{figspin1} and Fig.~\ref{figspin2}. The indirect exchange interaction is dominated by the intraband contribution of the susceptibility in the conduction band where the Fermi level is located, and is thus governed by the spin states of the electrons near the Fermi surface.   
As discussed in Sec.~\ref{EffHamiltonian}, unlike the equilibrium spin texture which always lies on the $xy$-plane (Fig.~\ref{figspin2}), the electronic spin texture under illumination acquires an additional spin component along the $z$-direction (Fig.~\ref{figspin1}). When the Fermi surface is near the bottom of the band, the electrons at the Fermi surface have spins that are predominantly polarized along the $z$-direction, so the induced spin polarization that mediates the indirect exchange interaction is also predominantly $z$-polarized. Therefore the exchange interaction of the electron spins with $\mathbb{S}^{\text{L,R}}_z$ is strengthened while that with $\mathbb{S}^{\text{L,R}}_y$ is weakened. 
This causes the $zz$ component of the exchange coupling to be enhanced and the other components $yy$ and $yz$ to be suppressed. When we increase the Fermi energy away from the band edge, the spin texture at the Fermi surface becomes less polarized along the $z$-axis, and the aforementioned effects become weaker. This is reflected in the smaller changes in the envelopes of $I_{A}$, $I_{B}$ and $I_{D}$ from their equilibrium values, as depicted in Fig.~\ref{fig5}. 
In the absence of irradiation, Eqs.~\eqref{analyticalA},~\eqref{analyticalB}, and~\eqref{analyticalD} show that the RKKY interaction energies are identical among the $yy$, $zz$ and $yz$ components, except for a $\pi/2$ phase difference in the $yz$ exchange oscillation relative to the $yy$ and $zz$ exchange oscillations. The total RKKY interaction is the sum of the contributions that we evaluated in Figs.~\ref{fig1},~\ref{fig2} and~\ref{fig3}. As the driving strength increases, we notice a stark contrast between the irradiated and the equilibrium RKKY interaction. With increasing light strengths, the Ising (zz) component of the exchange interaction dominates the total exchange. This can be understood again from the electronic spin texture, which becomes more aligned along the $z$-direction as the driving field is increased. Hence, the impurity spins in the spin chains will tend to align out of plane in the parallel or antiparallel directions, while in the absence of light, these spins tend to align in the plane.

%

Finally, we conduct a detailed comparison of our results with the RKKY interaction in other topological materials both in equilibrium and under an external drive. In equilibrium, our results are consistent with the previously reported results for the RKKY interaction between parallel impurity spin chains on the surface of a 3D topological insulator~\cite{Mahmoud1}. We then compare our results with the RKKY interaction between defect lines in gapless graphene at  equilibrium~\cite{GrapheneLine}. Eq.~(19) in Ref.~\cite{GrapheneLine} reported a similar $y^{-3/2}$ power-law decay. However, their result is limited to the Heisenberg exchange interaction because graphene has negligible spin-orbit coupling. Our present findings go further to elucidate that the Dzyaloshinsky-Moriya interaction also exhibits a $y^{-3/2}$ dependence. Secondly, as Ref.~\cite{GrapheneLine} employed a tight-binding description of graphene, not all quantities were explicitly reported in the RKKY coupling in Ref.~\cite{GrapheneLine}, \textit{e.g.}, $\mathcal{Q}(E)$ and $\mathcal{A}^l$ in their Eq.~(19). In contrast, our low-energy continuum description makes it possible for us to obtain explicit analytic results Eq.~\eqref{analyticalA}-Eq.~\eqref{analyticalD} for the long-range behavior of the RKKY interaction. Our analytic results for both the spin chain case and the isolated impurities case [Eq.~\eqref{SPASingleA}-Eq.~\eqref{SPASingleD}] also clarify that the oscillation phase difference between the two cases is $\pi/4$, which was not reported in Ref.~\cite{GrapheneLine}. Next, we compare our results with the recently reported results for topological crystalline insulators (TCI) driven under similar conditions~\cite{FloqTCI2}. In contrast to the single Dirac cone in our case, the energy spectrum of TCI features two low-energy Dirac cones. The presence of these two valleys causes additional inter-valley oscillations, with a period determined by the distance between the two Dirac points, to be superimposed on the usual intra-valley oscillations that are governed by the Fermi level. This phenomenon is reminiscent of the RKKY oscillations in graphene in equilibrium~\cite{Graphene1,Graphene2,Graphene3} and under periodic driving~\cite{Ke1}. Unlike the graphene case though, the momentum space separation between the two Dirac cones in TCI is only a fraction of the Brillouin zone size, resulting in a much longer period of the inter-valley oscillations. This period is comparable with the period of the intra-valley oscillations, leading to strong interference between the two types of oscillations. The resulting beating pattern obscures the intra-valley oscillations, and consequently presents additional difficulties in practice in deciphering the irradiation effects on the RKKY interaction, which arise predominantly through the intra-valley mechanisms of photon-dressed bands and light-induced spin texture changes. Hence, optically driven RKKY interaction should be more readily achievable in the surface states of strong TI as presented in this work than in TCI.

\section{Conclusion}\label{Conclusion}
We have presented a theory for the indirect exchange interaction energy between two parallel chains of magnetic impurities on the surface of a 3D topological insulator, irradiated under off-resonant circularly polarized light. Our theory is based on the high-frequency expansion of the effective Floquet Hamiltonian, which captures the light-induced gap at the Dirac point and renormalized Fermi velocity of the topological insulator surface states. An exact closed-form analytic expression of the spin susceptibility tensor of the irradiated topological insulator surface states is obtained. Our analytical and numerical results of the exchange couplings reveal that increasing the driving strength of light extends the oscillation period due to an increasing light-induced dynamical gap. Moreover, we find that light irradiation generates distinct oscillation envelopes of the Heisenberg, Ising and Dzyaloshinskii-Moriya contributions to the indirect exchange energy, suppressing the Heisenberg and Dzyaloshinskii-Moriya contributions but strengthening the Ising contribution. The light-induced spin texture provides a clear physical picture to understand these distinct behaviors. 
Our work clarifies the interplay between light-induced effects and dimensionality of the spin configuration on the RKKY interaction in topological materials. Since the 
Dzyaloshinskii-Moriya interaction plays a crucial role in the emergence and stabilization of topological spin textures, our findings may suggest that further investigations into optically driven exchange interactions could offer a useful strategy for realizing tunable magnetic patterns in topological materials. 

\acknowledgments
This work was supported by the U.S. Department of Energy, Office of Science, Basic Energy Sciences under Early Career Award No. DE-SC0019326 (M.K and W-K.T.), and by the National Science Foundation via Grant No. DMR-2213429 (M.M.A.).

\appendix{}
\section{Calculation of spin susceptibility}\label{app1}
\subsection{$zz$ case}
There is a linear ultraviolet divergence occurring in the calculation of the $zz$ component of the susceptibility.  In order to avoid the divergence encountered in the integral, we derive the renormalized susceptibility $\chi^{zz}_{\rm ren}(q)$, which is given by subtracting the bare susceptiblility $\chi^{zz}(q)$ by the intrinsic susceptibility at zero momentum $\chi^{zz}(q=0,\mu=0)$ \cite{Garate, Mahmoud1}.
By carrying out the $k$ integral Eq.\eqref{sus2} reduces to:
\begin{eqnarray}
&&\chi^{zz}_{\rm ren}(q)=\frac{2\mu_B^2}{(2\pi\alpha)^2}\int_{0}^{1} dx [2\pi(\sqrt{\delta^2(x)+\Delta^2}-\Delta)\nonumber\\
&&\Theta(\delta^2(x)+\Delta^2-\mu^2)+\pi(|\mu|-\Delta)\Theta(\mu^2-
\delta^2(x)-\Delta^2)\nonumber\\
&&-\pi\Delta\Theta(\mu^2-
\delta^2(x)-\Delta^2)],
\end{eqnarray}
where $\Theta(x)$ is the Heaviside step function.

The integration range of this integral where the theta functions are nonzero is given by $x_1=1/2-\sqrt{1-4(\mu^2-\Delta^2)/(\alpha^2 q^2)}/2$ and $x_2=1/2+\sqrt{1-4(\mu^2-\Delta^2)/(\alpha^2 q^2)}/2$. Carrying out the integral we obtain:
\begin{eqnarray}
&&\chi^{zz}_{\rm ren}(q)=\nonumber
\\
&&\begin{cases}
          \frac{\mu_B^2}{2\pi\alpha^2}\Re\bigg[|\mu|-2\Delta+\frac{1}{2}(\alpha q+\frac{4\Delta^2}{\alpha q})\sin^{-1}{\gamma}\bigg]\nonumber\\
           \text{if } |\mu|> \Delta,\nonumber\\
          \\
        \mu_B^2\frac{\alpha q+\frac{4\Delta^2}{\alpha q}}{4\pi\alpha^2}\Re\bigg[\sin^{-1}\gamma\bigg]\nonumber\\
          \text{if}  |\mu|< \Delta.
    \end{cases}\nonumber\\
\end{eqnarray}
\subsection{$xx$ and $yy$ case}
Similarly to avoid the divergence, we renormalize the susceptibility by $\chi^{xx}_{\rm ren}(q)=\chi^{xx}(q)-\chi^{xx}(q=0,\mu=0)$. By carrying out the $k$ integral Eq.\eqref{sus2} reduces to:
\begin{eqnarray}
&&\chi^{xx}_{\rm ren}(q)=\frac{2\mu_B^2}{(2\pi\alpha)^2}\int_{0}^{1} dx [\frac{\pi \delta^2(x)\cos^2{\phi}}{\sqrt{\delta^2(x)+\Delta^2}}\nonumber\\
&&\times\Theta(\delta^2(x)+\Delta^2-\mu^2)].
\end{eqnarray}
Carrying out this integral in the range of nonzero value of the theta function we get:
\begin{eqnarray}
&&\chi^{xx}_{\rm ren}(q)=\nonumber\\
&&\begin{cases}
          \frac{\mu_B^2}{4\pi\alpha^2}\Re\bigg[\sqrt{1-4\frac{\mu^2-\Delta^2}{\alpha^2 q^2}}|\mu|\\+\frac{1}{2}(\alpha q-\frac{4\Delta^2}{\alpha q})\sin^{-1}\gamma\bigg]\cos^2{\phi},& \text{if } |\mu|> \Delta\\
          \\
        \frac{\mu_B^2}{2\pi\alpha^2}\cos^2{\phi}\bigg[\frac{1}{2}\Delta+\alpha q \tan^{-1}{\frac{\alpha q}{2\Delta}}\\-\frac{1}{8}(3\alpha q+\frac{4\Delta^2}{\alpha q})(\pi-\sin^{-1}{\gamma})\bigg], & \text{if }  |\mu|< \Delta
    \end{cases}
\end{eqnarray}
The integral of the $yy$ component can be deduced very similarly.
\subsection{$xy$ and $yx$ case}
The integral of $xy$ and $yx$ component of the susceptibility does not diverge, and the renormalization does not change the result as $\chi^{xy}(q=0,\mu=0)=\chi^{yx}(q=0,\mu=0)=0$ in this case. The result of the $k$ integral gives:
\begin{eqnarray}
&&\chi^{xy}(q)=\chi^{yx}(q)=\frac{2\mu_B^2}{(2\pi\alpha)^2}\nonumber\\
&&\times\int_{0}^{1} dx\int dk_0 \frac{\delta^2(x)\sin{2\phi}}{\delta^2(x)+\Delta^2+(k_0-i\mu)^2}.
\end{eqnarray}
This integral takes the same form as in the $xx$ and $yy$ case, thus we can make use our previous result and obtain the following result:
\begin{eqnarray}
&&\chi^{xy}(q)=\chi^{yx}(q)=\nonumber\\
&&\begin{cases}
          \frac{\mu_B^2}{4\pi\alpha^2}\Re\bigg[\sqrt{1-4\frac{\mu^2-\Delta^2}{\alpha^2 q^2}}\\+\frac{1}{2}(\alpha q-\frac{4\Delta^2}{\alpha q})\sin^{-1}\gamma\bigg]\sin{2\phi},& \text{if } |\mu|> \Delta\\
          \\
        \frac{\mu_B^2}{2\pi\alpha^2}\sin{2\phi}\bigg[\frac{1}{2}\Delta+\alpha q \tan^{-1}{\frac{\alpha q}{2\Delta}}\\-\frac{1}{8}(3\alpha q+\frac{4\Delta^2}{\alpha q})(\pi-\sin^{-1}\gamma)\bigg] ,& \text{if }  |\mu|< \Delta
    \end{cases}
\end{eqnarray}
\begin{figure}\label{mag2}
   \begin{center}
            \includegraphics[width=1\columnwidth]{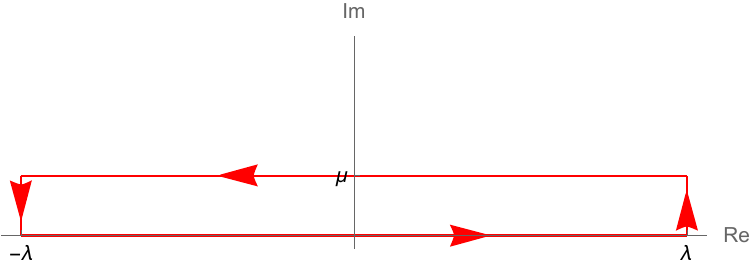}
                \end{center}
                \caption{The contour of the integral used in the calculation of $xz$, $zx$ $yz$ and $zy$ cases.}\label{fig6}
\end{figure}
\subsection{$xz$, $zx$ $yz$ and $zy$ case}
Here we show the derivation for the $xz$ case for example, and the other three cases follow the same procedure.
\begin{eqnarray}
\chi^{xz}(q)&=&\mu_B^2\frac{q}{(2\pi)^2\alpha}\int_{0}^{1} dx\int dk_0 \nonumber\\
&\times&\frac{(k_0-i\mu)\cos{\phi}+\Delta(2x-1)\sin{\phi}}{(k_0-i\mu)^2+\delta^2(x)+\Delta^2}.
\end{eqnarray}
We consider a rectangular contour in the complex plane of $k_0$ enclosed by the real line and a straight line with distance $i\mu$ above the real line as shown in Fig.~\ref{fig6}, and use the residue theorem to evaluate the integral. The singular point of the integrand is at $k_0=i\mu-i\sqrt{\delta^2(x)+\Delta^2}$. The contour consists of the upper and lower infinite integrals with the narrow-ranged integral on the two sides. It can be shown that the two side integrals are zero by evaluating the side integrals at a finite position $\pm\lambda$ on the real axis and take the limit of $\lambda\rightarrow\infty$. The upper integral is:
\begin{eqnarray}
I=\int_{0}^{1}\frac{dx}{(2\pi)^2}\frac{\pi q}{\alpha}\frac{\Delta(2x-1)\sin{\phi}}{\sqrt{\delta^2(x)+\Delta^2}}.
\end{eqnarray}
Subtracting this term from the residue and then carrying out the $x$ integral we get the susceptibility we need:
\begin{eqnarray}
&&\chi^{xz}(q)=\nonumber\\
&&\begin{cases}
          \frac{i q}{4\pi\alpha}\mu_B^2\bigg\{1-\Re\bigg[\sqrt{1-4\frac{\mu^2-\Delta^2}{\alpha^2 q^2}}\bigg]\bigg\}\cos{\phi}\\
           \text{if } |\mu|> \Delta,\\
          \\
       0\\
       \text{if }  |\mu|< \Delta.
    \end{cases}
\end{eqnarray}
Note that for some of the cases the integral contour needs to be set below the real axis, but the procedures are the same.

.

\section{A stationary phase approximation approach to the RKKY interaction}\label{app3}

In this appendix, we present an alternative derivation of the long-range asymptotic behavior of the RKKY interaction between two impurity spin chains on an irradiated TI surface, making use of the stationary phase approximation of the real-space Green's functions. Toward this end, we first obtain the RKKY interaction between two isolated impurities. 

In the momentum space, we can write the Green's function in the form of a spectral decomposition,
\begin{eqnarray}
&&G_0^\text{R}(\boldsymbol{k},E)=\label{GreenSpectral}\\
&&G_+^\text{R}(\boldsymbol{k},E)\ket{+,\boldsymbol{k}}\bra{\boldsymbol{k},+}+G_-^\text{R}(\boldsymbol{k},E)\ket{-,\boldsymbol{k}}\bra{\boldsymbol{k},-}, \nonumber
\end{eqnarray}
where $G_{\pm}^\text{R}(\boldsymbol{k},E)=
(E+i\eta-E_{k,\pm})^{-1}$ is the
the retarded Green's function for the conduction ($+$) and valence ($-$) bands, and $E_{k,\pm} = \pm \sqrt{\alpha^2k^2+\Delta^2}$ are their band energies. 
The real-space Green's function then follows from Eq.~\eqref{GreenSpectral} as 
\begin{eqnarray}
&&G_0^\text{R}(\boldsymbol{R},E) =\int \frac{d\boldsymbol{k}}{(2\pi)^2}e^{i\boldsymbol{k}\cdot\boldsymbol{R}}\int_{0}^{\infty}\frac{dt}{i\hbar}e^{i(E+i\eta)t/\hbar}\label{GreenSpectral2}
\\
&&\times\left[e^{-iE_{k,+}t/\hbar}\ket{+,\boldsymbol{k}}\bra{\boldsymbol{k},+}+e^{-iE_{k,-}t/\hbar}\ket{-,\boldsymbol{k}}\bra{\boldsymbol{k},-}\right].\nonumber
\end{eqnarray}
Following the standard procedures of SPA~\cite{Bender}, we find the stationary point of the phase function for each of the exponentials in Eq.~\eqref{GreenSpectral2}, make a Taylor expansion of those phases around their respective stationary points, and evaluate the resulting integrals. The real-space Green's function within SPA is then obtained as  
\begin{eqnarray}
G_0^\text{R}(\boldsymbol{R},E) &=& -\Theta(E^2-\Delta^2)\frac{i|E|}{\sqrt{2\pi R}\alpha^{\frac{3}{2}}(E^2-\Delta^2)^{\frac{1}{4}}}\tilde{G}  \nonumber
\\
&&\times \mathrm{exp}\left[i\mathrm{sgn}(E)\left(\frac{\sqrt{E^2-\Delta^2}R}{\alpha}-\frac{\pi}{4}\right)\right], \nonumber \\
\label{GreenSPA}
\end{eqnarray}
where $\Theta(x)$ denotes the Heaviside step function and
\begin{eqnarray}\label{SPAvector}
&&\tilde{G}=\begin{bmatrix}
    \frac{1}{2}(1+\frac{\Delta}{E}) & \frac{i}{2}\sqrt{1-\frac{\Delta^2}{E^2}}e^{-i\phi_R}\\
    \frac{i}{2}\sqrt{1-\frac{\Delta^2}{E^2}}e^{i\phi_R} &  \frac{1}{2}(1-\frac{\Delta}{E})
\end{bmatrix},
\end{eqnarray}
with $\phi_R$ being the azimuthal angle of $\mathbf{R}$. With the SPA Green's function, we next proceed to calculate the RKKY interaction between two single impurities. The exchange interaction energy can be obtained from the real-space Green's function as follows~\cite{RKKYsoc,Zhu,Dugaev},
\begin{eqnarray}\label{exchansingSPA1}
\tilde{I}=-\frac{1}{\pi}(\mu_BJ_0)^2\textrm{Im}\bigg\{\int_{-\infty}^{\mu}\,dE \textrm{Tr}
\big[(\boldsymbol{\sigma}\cdot\mathbb{S}^{\text{R}}) G_0^\text{R}(\boldsymbol{R},E)\nonumber
\\(\boldsymbol{\sigma}\cdot\mathbb{S}^{\text{L}})G_0^\text{R}(\boldsymbol{-R},E)\big] \bigg\}. 
\end{eqnarray}
Plugging in Eq.~(\ref{GreenSPA}) and performing integration by parts to obtain the dominant asymptotic behavior, we arrive at the following leading-order contributions to the exchange interaction:
\begin{equation}\label{SPASingleA}
\tilde{I}_A(R)\approx-\tilde{I}_{A,0}\frac{c}{\alpha_0}\frac{1}{|\mu|}\frac{\mu^2-\Delta^2}{(1-\frac{\Delta}{\hbar\Omega})^2}\left(\frac{c}{R}\right)^2\cos{\left(q_0 R-\frac{\pi}{2}\right)},
\end{equation}
\begin{equation}\label{SPASingleB}
\tilde I_B(R)\approx-\tilde{I}_{B,0}\frac{c|\mu|}{\alpha_0}\frac{1}{(1-\frac{\Delta}{\hbar\Omega})^2}\left(\frac{c}{R}\right)^2\cos{\left(q_0 R-\frac{\pi}{2}\right)},
\end{equation}
\begin{equation}
\tilde I_D(R)\approx-\tilde{I}_{D,0}\bigg[\frac{c}{\alpha_0}\frac{\sqrt{\mu^2-\Delta^2}}{(1-\frac{\Delta}{\hbar\Omega})^2}\left(\frac{c}{R}\right)^2\cos{\left(q_0 R-\pi\right)}\bigg],
\label{SPASingleD}
\end{equation}
where $\tilde{I}_{A,0}=\tilde{\mathcal{I}}\mathbb{S}^{\text{L}}_y\mathbb{S}^{\text{R}}_y$, $ \tilde{I}_{B,0}=\tilde{\mathcal{I}}\mathbb{S}^{\text{L}}_z\mathbb{S}^{\text{R}}_z$, $\tilde{I}_{D,0}=\tilde{\mathcal{I}}(\boldsymbol{\mathbb S}^{\text{L}}\times\mathbb S^{\text{R}})_x$, and $\tilde{\mathcal{I}}= [\mu_BJ_0/(2\pi c\sqrt{\alpha_0 c})]^2$.
We note that the explicit form of the asymptotic behavior of the RKKY interaction between isolated impurities on a magnetically gapped TI surface states has not been previously reported in the literature \cite{RKKYtopo}. We have also checked that the above results Eqs.~\eqref{SPASingleA}-\eqref{SPASingleD} can be obtained using the approximation method described in the main text that is based on the spin susceptibilities Eqs.~\eqref{sus3}-\eqref{sus6}.  

With the above results for the single impurity case, we can proceed to calculate the exchange interaction between two parallel impurity spin chains each with $N$ impurities and length $L$. As in our setup discussed in the main text, the chains are infinitely extended along the $x$-direction and separated in the $y$-direction, therefore $N,L \to \infty$ while the impurity density $N/L$ is finite. By summing up the contributions from each pair of impurity spins, the total interaction energy between the two spin chains is given by 
\begin{eqnarray}
I^{\text{tot}}=\sum_{m,n = -\infty}^{\infty} \tilde{I}(\vert \boldsymbol{R}_n-\boldsymbol{R}_m\vert),
\end{eqnarray}
where $\tilde{I}=\tilde I_A+\tilde I_B+\tilde I_D$ is the interaction energy between the $m^{\mathrm{th}}$ impurity in chain $\text{L}$ and the $n^{\mathrm{th}}$ impurity in chain $\text{R}$, $\boldsymbol{R}_{m}$ and $\boldsymbol{R}_{n}$ are the position vectors of  impurity $m \in \text{L}$ and impurity $n \in \text{R}$, respectively. Due to discrete translational symmetry along the $x$-direction, the  total interaction energy $I^{\text{tot}}$ is the same as $N$ times the interaction energy between a single impurity in chain $\text{L}$ and all impurities in chain $\text{R}$:
\begin{eqnarray}\label{sum_Itot}
I^{\text{tot}}=N\sum_{n = -\infty}^{\infty} \tilde{I}(R_n),
\end{eqnarray}
where $R_n=\sqrt{(n\Delta x)^2+y^2}$ with $\Delta x=2A_0/c$ being the distance between two adjacent impurities on one chain. The interaction energy per unit length $I$ is then given by $I^{\text{tot}}/L$. To obtain the leading asymptotic behavior, we first use the Euler–Maclaurin formula to approximate the sum in Eq.~\eqref{sum_Itot} as an integral, 
\begin{equation}
I=\frac{N}{L}\sum_{n = -\infty}^{\infty} \tilde{I}(R_n)\approx\left(\frac{N}{L}\right)^2\int_{-\infty}^{\infty} du \tilde{I}(\sqrt{u^2+y^2}),
\end{equation}
and evaluate the resulting integral within the SPA. We then recover the same results up to the leading order in $1/y$, Eqs.~(\ref{analyticalA})-(\ref{analyticalD}), obtained through the spin susceptibilities in the main text. 

\bibliography{Submissionfile}
\end{document}